\documentclass[11pt,epsf]{article}

\usepackage{epsfig}
\usepackage{amssymb}
\usepackage{graphicx}
\usepackage{color}
\usepackage{subfigure}
\usepackage{mathtools}

\makeatletter

\@addtoreset{equation}{section} \makeatother
\def\a{\alpha}
\def\b{\beta}
\def\g{\gamma}

\def\D{\Delta}
\def\ep{\epsilon}
\def\e{\eta}
\def\ph{\phi}
\def\Ph{\Phi}

\def\L{\Lambda}
\def\m{\mu}
\def\n{\nu}
\def\th{\theta}

\def\r{\rho}
\def\s{\sigma}

\def\ta{\tau}
\def\o{\omega}
\def\O{\Omega}
\def\pr{\prime}

\def\w{\wedge}

\def\p{\partial}

\def\lt{\left}
\def\rt{\right}

\def\nn{\nonumber}

\DeclarePairedDelimiter\abs{\lvert}{\rvert}%

\begin{document}

\begin{titlepage}
\title{\vskip -60pt
\vskip 20pt $\b$-deformation on a slanted torus and deformed pp-wave
}
\author{
Ariunzul Dagvadorj 
\footnote{e-mail : ariunzul.d@gmail.com},
Sunyoung Shin\footnote{e-mail : sihnsy@gmail.com}}
\date{}
\maketitle \vspace{-1.0cm}
\begin{center}
~~~
\it Bogoliubov Laboratory of Theoretical Physics, JINR, 141980
Dubna, Moscow region, Russia
~~~\\
~~~\\
\end{center}

\begin{abstract}
We discuss the $\b$-deformation of $AdS_5\times S^5$ which
incorporates the $SL(2,\mathbb{R})$ symmetry of the type IIB theory.
The axion-dilaton is identified with a two-torus from an eleven
dimensional viewpoint. We consider the null geodesic with equal
component angular momenta to take the Penrose limit of the deformed
$AdS_5\times S^5$. We study the bosonic part of the string sigma
model and the spectrum of the string in the pp-wave background.
\end{abstract}

\end{titlepage}
\newpage


\section{Introduction}
\setcounter{equation}{0}
%
The marginal deformation of $\mathcal{N}=4$ super Yang-Mills theory
introduces phases in the superpotential, preserving a $U(1)\times
U(1)$ non-R-symmetry. The deformation reduces the supersymmetry
$\mathcal{N}=4$ to $\mathcal{N}=1$. The phases in the superpotential
can be complexified\footnote{We use $\g$ for the real parameter and
$\b$ for the complex parameter.}\cite{Leigh:1995ep}. In the gravity
side, the $U(1)\times U(1)$ symmetry maps to a two-torus. An
$SL(2,\mathbb{R})$ transformation acting on a type IIB supergravity
solution compactified on the two-torus produces the gravity dual of
the $\g$-deformation. The gravity dual of the $\b$-deformation is an
$SL(3,\mathbb{R})$ transformation, which consists of the
$SL(2,\mathbb{R})$ transformation and an S-duality transformation
$SL(2,\mathbb{R})_s$\footnote{We use $\s$ for the S-duality
transformation.}\cite{Lunin:2005jy}.

The charges of the chiral superfields under $U(1)\times U(1)$ in the
gauge theory correspond to the angular momenta along the two-torus.
In the marginally deformed $AdS_5\times S^5$, the $SO(6)$ isometry
is broken to $U(1)\times U(1)\times U(1)$. The angle coordinates
$(\phi_1,\phi_2,\phi_3)$ of the $S^5$ are linear combinations of the
coordinates of the $U(1)\times U(1)\times U(1)$. The BPS geodesics
are chosen with angular momenta
\begin{eqnarray}\label{eq:js}
(J_{\phi_1},J_{\phi_2},J_{\phi_3})\sim(J,0,0),~(0,J,0),~(0,0,J),~(J,J,J).
\end{eqnarray}
For undeformed $AdS_5\times S^5$, the geodesics can be transformed
to one another by $SO(6)$, which is the isometry of the $S^5$.
Therefore the Penrose limits for the geodesics produce one pp-wave.
For deformed $AdS_5\times S^5$, which has a $U(1)\times U(1)\times
U(1)$ symmetry, the geodesics are not isometrically equivalent. The
first three geodesics and the fourth geodesic are two distinct
geodesics. The Penrose limit for the first three cases is studied in
\cite{Niarchos:2002fc,Lunin:2005jy}. The Penrose limit for the
fourth case is studied in \cite{deMelloKoch:2005vq} where it is also
shown that the spectrum of the string in this pp-wave limit is
independent of the parameter $\g$. The pp-wave limits of marginally
deformed geometries which include the $\s$-deformation are discussed
in \cite{Avramis:2007wb}. Giant gravitons on the deformed pp-waves
are investigated in \cite{Avramis:2007wb,Hamilton:2006ri}.

The $SL(3,\mathbb{R})$ transformation for the $\b$-deformation can
be generalized by incorporating the $SL(2,\mathbb{R})$ symmetry of
type IIB theory, which is also the symmetry of the toroidal
compactification \cite{Aspinwall:1995fw}. In \cite{Shin:2013oya},
torus deformation is considered for the generalization. In this
work, we apply the generalized $\b$-deformation to $AdS_5\times S^5$
and take the Penrose limit of the deformed $AdS_5\times S^5$ along
the $(J,J,J)$ geodesic. We study the spectrum of the string in the
deformed pp-wave background.

In section \ref{sec:deform}, we review the generalization of the
$\b$-deformation and present the deformed $AdS_5\times S^5$
geometry. In section \ref{sec:ppwave}, we study the pp-wave limit of
the $\b$-deformed $AdS_5\times S^5$ with the axion-dilaton, which is
identified with the torus modulus of the rectangular torus before
the torus deformation. We present the bosonic part of the string
sigma model and compute the spectrum. In section \ref{sec:discuss},
we summarize our results.

\section{Generalization of the $\b$-deformed geometry}\label{sec:deform}
\setcounter{equation}{0}
The $\b$-deformation \cite{Lunin:2005jy} acting on a type IIB
supergravity solution which has a two-torus symmetry is derived from
an $SL(3,\mathbb{R})$ transformation acting on an eleven-dimensional
supergravity solution which has a three-torus symmetry. The
coordinates of the three-torus are
$(\varphi_1,\varphi_2,\varphi_3)$. The type IIB supergravity
solution is obtained by a dimensional reduction along $\varphi_3$
and a T-duality transformation along $\varphi_1$. The
$SL(3,\mathbb{R})$ matrix for the $\b$-deformation is
\begin{eqnarray}\label{eq:LM}
\L^T_{LM}=\lt(\begin{array}{ccc}
1   &   0   &  0 \\
\g         &   1   &  \s       \\
0   &   0   &  1
\end{array}\rt).
\end{eqnarray}
The type IIB supergravity solution can be generalized by an
$SL(3,\mathbb{R})$ transformation
\begin{eqnarray}\label{eq:t3def}
&&L=\lt(\begin{array}{ccc}
L_{11}   &   0   &  L_{13} \\
0         &   1   &  0      \\
L_{31}   &   0   &  L_{33}
\end{array}\rt),~~L_{11}L_{33}-L_{13}L_{31}=1.
\end{eqnarray}
This corresponds to the $SL(2,\mathbb{R})$ symmetry of the type IIB
theory, which is also the symmetry of the toroidal compactification
\cite{Aspinwall:1995fw}. The axion-dilaton $\ta=\ta_1+i\ta_2$
transforms as
\begin{eqnarray}\label{eq:sl2_frac}
\ta^\prime=\frac{L_{11}\ta+L_{31}}{L_{13}\ta+L_{33}}.
\end{eqnarray}
The $SL(3,\mathbb{R})$ transformation $L\L^T_{LM}$, therefore,
produces the $\b$-deformed geometry incorporating the
$SL(2,\mathbb{R})$ symmetry of the type IIB supergravity. By
applying Lunin and Maldacena's solution generating technique with
the $SL(3,\mathbb{R})$ matrix $L\L^T_{LM}$ to the type IIB
supergravity solution in the form given by (A.7) in
\cite{Lunin:2005jy}, the generalized $\b$-deformation is obtained in
\cite{Shin:2013oya}\footnote{We follow the formulas and the
conventions of \cite{Lunin:2005jy}. It is assumed that only the
metric, the complex field $\chi_0+ie^{-\Phi_0}=\ta=\ta_1+i\ta_2$ and
$\tilde{d}_{\m\n}$ are excited and the other fields are zero.} as
\begin{eqnarray}\label{eq:gen_beta}
&&ds^{\pr
2}=F^\pr\Big[\frac{1}{\sqrt{\D}}(D\varphi_1-CD\varphi_2)^2+\sqrt{\D}(D\varphi_2)^2\Big]+\frac{e^{2\Phi^\pr/3}}{F^{\pr1/3}}g_{\m\n}dx^\m
dx^\n, \nn\\
&&F^\pr=FG\sqrt{H},~e^{\Phi^\pr}=\sqrt{G}H\ta_2^{-1},~\chi^\pr=H^{-1}(h+\g\s\ta_2^2F^2),\nn\\
&&B_2^\pr=GF^2(\g f-\s h)D\varphi_1\w
D\varphi_2+\frac{\s}{2}\tilde{d}_{\m\n}dx^\m\w dx^\n,\nn\\
&&C_2^\pr=GF^2(\g h-\s g)D\varphi_1\w
D\varphi_2+\frac{\g}{2}\tilde{d}_{\m\n}dx^\m\w dx^\n,\nn\\
&&F_5^\pr=\tilde{F}_5+\star \tilde{F}_5,
\end{eqnarray}
where
\begin{eqnarray}
G^{-1}&=&1+(\g^2 f-2\g\s h+\s^2 g)F^2,\nn\\
H&=&f+\ta_2^2\s^2 F^2,
\end{eqnarray}
\begin{eqnarray}\label{eq:fgh}
f&=&\lt(L_{33}+L_{13}\ta_1\rt)^2+{L_{13}}^2 \ta_2^2, \nn\\
g&=&\lt(L_{31}+L_{11}\ta_1\rt)^2+{L_{11}}^2 \ta_2^2, \nn\\
h&=&\lt(L_{33}+L_{13}\ta_1\rt)\lt(L_{31}+L_{11}\ta_1\rt)+L_{11}L_{13}
\ta_2^2,
\end{eqnarray}
$\tilde{F}_5$ in (\ref{eq:gen_beta}) has no indices along the torus
$(\varphi_1,\varphi_2)$. The star is taken with the new metric.

We consider $AdS_5\times S^5$ defined by
\begin{eqnarray}\label{eq:ads}
&&ds^2=R^2\Big[-dt^2\cosh^2\r+d\r^2+\sinh^2\r d\O_3^2+\sum_{i=1}^3d\m_i^2+\sum_{i=1}^3\m_i^2d\ph_i^2\Big], \nn\\
&&e^{-\Phi_0}=\ta_2,~\chi_0=\ta_1,~B_2=0,~C_2=0,\nn\\
&&C_4=4R^4\ta_2(\o_4+\o_1\w d\ph_1\w d\ph_2\w d\ph_3),\nn\\
&&F_5=4R^4\ta_2(\o_{AdS_5}+\o_{S^5}),\nn\\
&&\o_{AdS_5}=d\o_4,~\o_{S^5}=d\o_1\w d\ph_1\w d\ph_2\w d\ph_3,\nn\\
&&d\o_1=\cos\a\sin^3\a\cos\th\sin\th d\a\w d\th,
\end{eqnarray}
with
\begin{eqnarray}\label{eq;beta_ads2}
&&\m_1=\cos\a,~\m_2=\sin\a\cos\th,~\m_3=\sin\a\sin\th.
\end{eqnarray}
$R$ is the radius of $AdS_5$ and the radius of $S^5$. We apply the
transformation (\ref{eq:gen_beta}) to (\ref{eq:ads}). The angle
coordinates $(\phi_1,\phi_2,\phi_3)$ are related to the coordinates
$(\varphi_1,\varphi_2)$ of the two-torus and the $U(1)$ R-symmetry
direction $\psi$ as
\begin{eqnarray}
\ph_1=\psi-\varphi_2,~\ph_2=\psi+\varphi_1+\varphi_2,~\ph_3=\psi-\varphi_1.
\end{eqnarray}
The deformed $AdS_5\times S^5$ geometry with parameters $\hat{\g}=\g
R^2$ and $\hat{\s}=\s R^2$ is
\begin{eqnarray} \label{eq:beta_ads1}
ds^{\pr\,2}&=&R^2H^{1/2}\Big[-dt^2\cosh^2\r+d\r^2+\sinh^2\r d\O_3^2\nn\\
&&+\sum_{i=1}^3\lt(d\m_i^2+G\m_i^2d\ph_i^2\rt)
+G\lt(\hat{\g}^2f-2\hat{\g}\hat{\s}h+\hat{\s}^2g\rt)\m_1^2\m_2^2\m_3^2\lt(\sum_{i=1}^3d\ph_i\rt)^2\Big],
\nn\\
e^{\Phi^\pr}&=&\sqrt{G}H\ta_2^{-1},\nn\\
\chi^\pr&=&H^{-1}\lt(h+\ta_2^2\hat{\g}\hat{\s}g_0\rt),\nn\\
B_2^\pr&=&R^2G\lt(\hat{\g}f-\hat{\s}h\rt)\o_2-4R^2\ta_2\hat{\s}\o_1\w\sum_{i=1}^3d\ph_i,
\nn\\
C_2^\pr&=&R^2G\lt(\hat{\g}h-\hat{\s}g\rt)\o_2-4R^2\ta_2\hat{\g}\o_1\w\sum_{i=1}^3d\ph_i,\nn\\
C_4^\pr&=&4R^4\ta_2\o_4+4R^4\ta_2G\Big[1-(\hat{\g}\hat{\s}h-\hat{\s}^2g)g_0\Big]\o_1\w
d\ph_1\w d\ph_2\w d\ph_3,\nn\\
F_5^\pr&=&4R^4\ta_2(\o_{AdS_5}+G\o_{S^5}),
\end{eqnarray}
where
\begin{eqnarray}
G^{-1}&=&1+\lt(\hat{\g}^2f-2\hat{\g}\hat{\s}h+\hat{\s}^2g\rt)g_0,
\nn\\
H&=&f+\ta_2^2\hat{\s}^2g_0, \nn\\
g_0&=&\m_1^2\m_2^2+\m_2^2\m_3^2+\m_3^2\m_1^2,\nn\\
\o_2&=&\m_1^2\m_2^2d\ph_1\w d\ph_2+\m_2^2\m_3^2d\ph_2\w
d\ph_3+\m_3^2\m_1^2d\ph_3\w d\ph_1.
\end{eqnarray}
$f$, $g$ and $h$ are the same as (\ref{eq:fgh}).

The transformation (\ref{eq:t3def}) can be related to torus
deformation from an eleven dimensional viewpoint. The torus
parameters considered in \cite{Shin:2013oya} are
\begin{eqnarray}\label{eq:def_comp}
L_{11}=1,~L_{13}=\frac{r_3}{R_1}\cos\xi,~L_{31}=0,~L_{33}=1,
\end{eqnarray}
where $R_i,~(i=1,3)$ are the radii of the torus before the
deformation and $r_3=\frac{R_3}{\sin\xi}$ is the radius of the third
direction after the deformation. $\xi$ is the intersection angle
between the direction along the first coordinate and the direction
along the third coordinate of the slanted torus deformed by
(\ref{eq:t3def}) with the components (\ref{eq:def_comp}). We
consider a simpler case in which the axion-dilaton is identified
with the torus modulus of the rectangular torus before the
deformation \cite{Shin:2013oya} as
\begin{eqnarray}\label{eq:ta0}
\ta=\ta_1+i\ta_2=i\frac{R_1}{R_3}=il.
\end{eqnarray}
The axion-dilaton (\ref{eq:ta0}) transforms under
(\ref{eq:sl2_frac}) with the components (\ref{eq:def_comp}) as
\begin{eqnarray}\label{eq:tau}
\ta^\prime=\frac{R_1}{r_3}e^{i\xi}=l(\sin\xi\cos\xi+i\sin^2\xi).
\end{eqnarray}
This is the torus moduli of the deformed torus. By substituting
(\ref{eq:ta0}) into (\ref{eq:beta_ads1}) we find the $\b$-deformed
$AdS_5\times S^5$ on the slanted torus
\begin{eqnarray}\label{eq:simp_def}
ds^2&=&R^2\widetilde{H}^{1/2}\Big[-dt^2\cosh^2\r+d\r^2+\sinh^2\r
d\O_3^2+\sum_{i=1}^3\lt(d\m_i^2+\widetilde{G}\m_i^2d\ph_i^2\rt)
+9\widetilde{G}{\mathcal{P}}\m_1^2\m_2^2\m_3^2d\psi^2\Big],\nn\\
e^{\Ph^\pr}&=&\sqrt{\widetilde{G}}\widetilde{H}l^{-1},\nn\\
\chi^\pr&=&\widetilde{H}^{-1}\lt(l\cot\xi+\hat{\g}\hat{\s}l^2g_0\rt),\nn\\
B_2^\pr&=&R^2\widetilde{G}{\mathcal{Q}}\o_2-12R^2\hat{\s}l\o_1\w
d\psi,\nn\\
C_2^\pr&=&R^2\widetilde{G}\mathcal{T}\o_2-12R^2\hat{\g}l\o_1\w
d\psi,\nn\\
C_4^\pr&=&4R^4l\o_4+4R^4l\tilde{G}(1-\mathcal{U}g_0)\o_1\w d\ph_1\w
d\ph_2\w d\ph_3,\nn\\
F_5^\pr&=&4R^4l(\o_{AdS_5}+\widetilde{G}\o_{S^5}),
\end{eqnarray}
where
\begin{eqnarray}
\widetilde{G}^{-1}&=&1+{\mathcal{P}}g_0,\nn\\
\widetilde{H}&=&\csc^2\xi+\hat{\s}^2l^2g_0, \nn\\
{\mathcal{P}}&=&\hat{\g}^2\csc^2\xi-2\hat{\g}\hat{\s}l\cot\xi+\hat{\s}^2l^2,
\nn\\
{\mathcal{Q}}&=&\hat{\g}\csc^2\xi-\hat{\s}l\cot\xi,\nn\\
{\mathcal{T}}&=&\hat{\g}l\cot\xi-\hat{\s}l^2,\nn\\
{\mathcal{U}}&=&\hat{\g}\hat{\s}l\cot\xi-\hat{\s}^2l^2.
\end{eqnarray}
This geometry contains four parameters. The parameters $\hat{\g}$
and $\hat{\s}$ arise from the marginal deformation. The parameters
$l$ and $\xi$ arise from the axion-dilaton.

\section{$(J,J,J)$ geodesic }\label{sec:ppwave}
\setcounter{equation}{0}

We investigate the Penrose limit of the geometry (\ref{eq:simp_def})
along the geodesic with equal component angular momenta. This
corresponds to $(\m_1^2,\m_2^2,\m_3^2)=(1/3,1/3,1/3)$ in
(\ref{eq;beta_ads2}). In the vicinity of the geodesic with
$\a_0=\arccos({1}/{\sqrt{3}})$ and $\th_0={\pi}/{4}$, we set
\begin{eqnarray}
&&\a=\a_0-\frac{1}{\Xi^{1/4}R}x^2,~~
\th=\th_0+\sqrt{\frac{3}{2}}\frac{1}{\Xi^{1/4}R}x^1,~~
\r=\frac{1}{\Xi^{1/4}R}r,\nn\\
&&\varphi_1=\sqrt{\frac{3+{\mathcal{P}}}{2}}\frac{1}{\Xi^{1/4}R}\lt(x^3-\frac{1}{\sqrt{3}}x^4\rt),~~~
\varphi_2=\sqrt{\frac{2(3+{\mathcal{P}})}{3}}\frac{1}{\Xi^{1/4}R}x^4,\nn\\
&&t=x^++\frac{1}{({\Xi^{1/4}R})^2}x^-,~~~\psi=-x^++\frac{1}{({\Xi^{1/4}R})^2}x^-,\nn\\
&&\Xi=\csc^2\xi+\frac{1}{3}\hat{\s}^2l^2,
\end{eqnarray}
and take the $R\rightarrow\infty$ limit of the geometry keeping
$\hat{\g}$ and $\hat{\s}$ fixed. We also shift the coordinate $x^-$
as $x^-\rightarrow
x^-+\frac{\sqrt{3}}{2\sqrt{3+\mathcal{P}}}\lt(x^1x^3+x^2x^4\rt)$ to
transform the resulting metric to a homogeneous pp-wave
\cite{Blau:2001ne,Papadopoulos:2002bg,Metsaev:2001bj}.

The bosonic part of the string sigma model is
\begin{eqnarray}\label{eq:st_sigma}
S=-\frac{1}{4\pi\a^\pr}\int d\ta
d\s\lt[\sqrt{\e}\e^{ab}g_{\m\n}\p_aX^\m\p_bX^\n+\ep^{ab}B_{\m\n}\p_aX^\m\p_bX^\n\rt],
\end{eqnarray}
where $\a^\pr=1/2\pi$, $0\leq\s\leq\pi$ and the worldsheet metric
$\e^{ab}$ is fixed as $\sqrt{\e}\e^{ab}=\mathrm{diagonal}(-1,1)$
with $\e=\abs{\mathrm{\det}\e_{ab}}$. We impose the lightcone gauge
condition $x^+=\ta$. The Lagrangian density of the action
(\ref{eq:st_sigma}) becomes
\begin{eqnarray}
{\mathcal{L}}&=&-2x^-_\ta-\frac{1}{2}\Big\{\sum_{i=1}^8\lt[-(x^i_\ta)^2+(x^i_\s)^2\rt]+\sum_{i=5}^8(x^i)^2
+\frac{4{\mathcal{P}}}{3+{\mathcal{P}}}\lt[(x^1)^2+(x^2)^2\rt]\Big\}\nn\\
&&+\frac{\sqrt{3}}{\sqrt{3+\mathcal{P}}}\Big[-x^3x^1_\ta-x^4x^2_\ta+x^1x^3_\ta+x^2x^4_\ta\Big]
+\frac{2\mathcal{Q}}{\sqrt{3+\mathcal{P}}}{\Xi}^{-1/2}\Big[x^2x^3_\s-x^1x^4_\s\Big]\nn\\
&&-12\hat{\s}l{\Xi}^{-1/2}\Big[\zeta
x^2x^1_\s-\lt(\frac{\sqrt{3}}{9}-\zeta\rt)x^1x^2_\s\Big],
\end{eqnarray}
where $x^i_\ta={\p x^i}/{\p \ta}$, $x^i_\s={\p x^i}/{\p \s}$.
$\zeta$ is a gauge parameter, which arises from $\o_1$ in
(\ref{eq;beta_ads2}). The $\zeta$ terms cancel out in the equations
of motion. We need to solve the equations of motion for $x^i$,
$(i=1,\cdots,4)$, which belong to the deformed $S^5$. The equations
of motion are
\begin{eqnarray}\label{eq:cond}
\frac{\p^2x^i}{\p\ta^2}-\frac{\p^2x^i}{\p\s^2}+f^{ij}\frac{\p
x^j}{\p\ta}+h^{ij}\frac{\p x^j}{\p\s}+k^ix^i=0,
\end{eqnarray}
with nonzero coefficients
\begin{eqnarray}
&&f^{13}=-f^{31}=f^{24}=-f^{42}=-\frac{2\sqrt{3}}{\sqrt{3+\mathcal{P}}},\nn\\
&&h^{12}=-h^{21}=-\frac{4}{\sqrt{3}}\hat{\s}l\Xi^{-1/2},\nn\\
&&h^{14}=-h^{41}=-h^{23}=h^{32}=\frac{2\mathcal{Q}\Xi^{-1/2}}{\sqrt{3+\mathcal{P}}},\nn\\
&&k^1=k^2=\frac{4\mathcal{P}}{3+\mathcal{P}}.
\end{eqnarray}
We solve the differential equations by the mode expansion
$x^i(\ta,\s)=\sum_{n=-\infty}^\infty
x_n^i(\ta)e^{2in\s},~x^i_n=(x_{-n}^i)^\ast$ with a harmonic
oscillator frequency ansatz $x_n^i(\ta)\sim u^i(\o_n)e^{i\o_n\ta}$
\cite{deMelloKoch:2005vq}. From the condition for the existence of
nontrivial solutions, we obtain the equation
\begin{eqnarray}\label{eq:omega}
\o^8+c_6\o^6+c_4\o^4+c_2\o^2+c_0=0,
\end{eqnarray}
with coefficients
\begin{eqnarray}
c_6&=&-8-16n^2,\nn\\
c_4&=&16+\frac{32}{3}\lt(6-\frac{\hat{\s}^2l^2}{\Xi}\rt)n^2+96n^4,\nn\\
c_2&=&-\frac{128}{3}\frac{\hat{\s}^2l^2}{\Xi}n^2-128\lt(1-\frac{2}{3}\frac{\hat{\s}^2l^2}{\Xi}\rt)n^4-256n^6,\nn\\
c_0&=&\frac{256}{9}\frac{\hat{\s}^4l^4}{\Xi^2}n^4-\frac{512}{3}\frac{\hat{\s}^2l^2}{\Xi}n^6+256n^8.
\end{eqnarray}
The solutions are
\begin{eqnarray}
\o=1\pm\sqrt{1+4n^2\pm\frac{4n\hat{\s}l}{\sqrt{3\csc^2\xi+\hat{\s}^2l^2}}}.
\end{eqnarray}
The spectrum does not depend on the deformation parameter $\hat{\g}$
while the spectrum depends on the deformation parameter $\hat{\s}$.
When $\hat{\s}\neq0$, the axion-dilaton parameters $l$ and $\xi$
contribute to the spectrum. When $\hat{\s}=0$, we recover the result
of \cite{deMelloKoch:2005vq}.

\section{Discussion}\label{sec:discuss}
We have applied the $\b$-deformation, which incorporates the
$SL(2,\mathbb{R})$ symmetry of the type IIB theory, to $AdS_5\times
S^5$. The $SL(2,\mathbb{R})$ parameters can be related to torus
parameters from an eleven dimensional viewpoint. The
$\b$-deformation becomes simpler when the axion-dilaton is
identified with the torus modulus of the rectangular torus before
the torus deformation. We have chosen the geodesic with equal
component angular momenta to take the Penrose limit of the
$\b$-deformed $AdS_5\times S^5$, which contains four parameters
arising from the marginal deformation and the axion-dilaton. We have
presented the string sigma model and obtained the spectrum of the
string in the deformed pp-wave limit. The spectrum does not depend
on $\hat{\g}$ while the spectrum depends on $\hat{\s}$. The
axion-dilaton parameters contribute to the spectrum when
$\hat{\s}\neq0$.

\vspace{1cm}


\end{document}